**The Surface Roughness of (433) Eros as Measured by Thermal-Infrared Beaming**

B. Rozitis[a]

[a]Planetary and Space Sciences, School of Physical Sciences, The Open University, Walton Hall, Milton Keynes, MK7 6AA, UK

No. of Manuscript Pages: 23
No. of Figures: 8
No. of Tables: 2

Please direct editorial correspondence and proofs to:

Dr. Ben Rozitis
Planetary and Space Sciences
School of Physical Sciences
The Open University
Walton Hall
Milton Keynes
Buckinghamshire
MK7 6AA
UK

Phone: +44 (0) 1908 332430
Fax: +44 (0) 1908 655667

Email: benjamin.rozitis@open.ac.uk

**ABSTRACT**

In planetary science, surface roughness is regarded to be a measure of surface irregularity at small spatial scales, and causes the thermal-infrared beaming effect (i.e. re-radiation of absorbed sunlight back towards to the Sun). Typically, surface roughness exhibits a degeneracy with thermal inertia when thermophysical models are fitted to disc-integrated thermal-infrared observations of asteroids because of this effect. In this work, it is demonstrated how surface roughness can be constrained for near-Earth asteroid (433) Eros (i.e. the target of NASA's NEAR Shoemaker mission) when using the Advanced Thermophysical Model with thermal-infrared observations taken during an "almost pole-on" illumination and viewing geometry. It is found that the surface roughness of (433) Eros is characterised by an RMS slope of 38 ± 8° at the 0.5-cm spatial scale associated with its thermal-infrared beaming effect. This is slightly greater than the RMS slope of 25 ± 5° implied by the NEAR Shoemaker laser ranging results when extrapolated to this spatial scale, and indicates that other surface shaping processes might operate, in addition to collisions and gravity, at spatial scales under one metre in order to make asteroid surfaces rougher. For other high obliquity asteroids observed during "pole-on" illumination conditions, the thermal-infrared beaming effect allows surface roughness to be constrained when the sub-solar latitude is greater than 60°, and if the asteroids are observed at phase angles of less than 40°. They will likely exhibit NEATM beaming parameters that are lower than expected for a typical asteroid at all phase angles up to 100°.

**Running head:**

Surface Roughness of (433) Eros

# 1. INTRODUCTION

## 1.1 Thermal Inertia and Surface Roughness

Thermal inertia and surface roughness are two useful properties for characterising an airless planetary surface. In particular, thermal inertia is a measure of a materials resistance to temperature change, and introduces a lag time between absorption and re-radiation of solar radiation. Increasing the thermal inertia decreases the day-side surface temperature distribution and increases it for the night-side of a planetary surface. Thermal inertia depends predominantly on the regolith particle size and depth, and on the degree of compaction and exposure of solid rocks and boulders within the top few decimetres of the surface (Delbo et al. 2007). Therefore, it can be used to infer the presence or absence of loose regolith material, and has been exploited in verifying the suitability of asteroid sample return targets [e.g. by Emery et al. (2014) for (101955) Bennu to be visited by OSIRIS-REx, and by Müller et al. (2011) for (162173) Ryugu to be visited by Hayabusa 2]. Thermal inertia also controls the strength of the asteroidal Yarkovsky effect (Vokrouhlický et al. 2015), and knowing its value can help measure the mass and density of an asteroid when a Yarkovsky orbital drift measurement is made (e.g. Rozitis et al. 2013, 2014; Rozitis and Green 2014; Chesley et al. 2014). It does not affect the YORP rotational acceleration of an asteroid (Čapek & Vokrouhlický 2004) but does influence the YORP effect on the asteroid obliquity (e.g. Statler 2015).

On the other hand, surface roughness is defined as the topographic expression of surfaces at horizontal scales of millimetres to a few hundred metres (Shepard et al. 2001). In terms of thermal models for planetary surfaces, it is regarded to be a measure of surface irregularity at scales smaller than the resolution of the shape model used but larger than the thermal skin depth specified by the thermal inertia (Rozitis & Green 2011). Surface roughness causes a planetary surface to thermally emit in a non-Lambertian way with a tendency to re-radiate the absorbed solar radiation back towards the Sun, an effect known as thermal–infrared beaming (Lagerros 1998). Beaming has been shown to be the result of two different processes: a rough surface will have elements orientated towards the Sun that become significantly hotter than a flat surface, and multiple scattering of radiation between rough surface elements increases the total amount of solar radiation absorbed by the surface (e.g. Rozitis & Green 2011; Davidsson et al. 2015). Contrary to expectation, the flux enhancement caused by beaming is greatest for limb surfaces rather than surfaces located at the sub-solar region (Rozitis & Green 2011). In addition for asteroids, surface roughness has been demonstrated to enhance the degree of Yarkovsky orbital drift, and to dampen, in general, the degree of YORP rotational acceleration (Rozitis & Green 2012).

Separating the effects of thermal inertia and surface roughness from disc-integrated thermal-infrared observations of asteroids is difficult because they both influence an asteroid's disc-integrated thermal spectrum in similar ways. Very commonly, a degeneracy between thermal inertia and surface roughness is obtained when a thermophysical model is fitted to thermal-infrared data of an asteroid [e.g. see Müller et al. (2012) when fitting to data of (101955) Bennu]. Typically, low surface roughness gives a lower bound on the thermal inertia value whilst high surface roughness gives an upper thermal inertia bound (e.g. see Figure 1), or vice-versa depending on the observational circumstances. However, it is possible to separately derive thermal inertia and surface roughness for spatially resolved bodies, as demonstrated by ground-based observations of the Moon [e.g. Rozitis & Green (2011) and references therein], and by in-situ spacecraft observations of asteroids [e.g. see Keihm et al. (2012, 2015) for asteroids (4) Vesta and (21) Lutetia] and comets [e.g. see Davidsson et al. (2013) and Groussin et al. (2013) for Comets 9P/Tempel 1 and 103P/Hartley 2].

In section 2 of this paper, it is demonstrated how disc-integrated thermal-infrared observations of asteroid (433) Eros can constrain its surface roughness when taken at an "almost pole-on" geometry via thermophysical modelling. This degree of surface roughness is then compared with that measured by other means during the NEAR Shoemaker mission in section 3. A summary with conclusions is provided in section 4.

## 1.2 Asteroid (433) Eros

(433) Eros is the second largest near-Earth asteroid (hereafter referred to as Eros), is an S-type asteroid, and was the target of NASA's NEAR Shoemaker mission (Cheng 2002). NEAR Shoemaker was the first spacecraft to orbit an asteroid when it arrived at Eros on 14[th] February 2000, and also became the first spacecraft to land on an asteroid when it made a controlled descent to the surface to end the mission on 12[th] February 2001. During its orbital phase, NEAR Shoemaker conducted a global mapping campaign that included several low altitude flybys. It obtained more than 160,000 multispectral images (MSI; Bussey et al. 2002), more than 200,000 spatially resolved near infrared reflectance spectra (NIS; Bell et al. 2002), and more than 16 million laser range measurements (NLR; Cheng et al. 2002). The MSI images had resolutions ranging from 3 to 19 m per pixel, which revealed an irregular shaped body with many craters and evidence of a surface regolith. The global shape model and rotational state of Eros was determined using feature tracking of specific targets easily visible on the asteroid surface (Thomas et al. 2002). This revealed a body with dimensions of 34.4 x 11.2 x 11.2 km (16.92 km mean diameter), a rotation period of 5.27 hr, and a rotation pole orientation at a high 89° obliquity. The images and continuity of the shape model also indicated that Eros was a solid body throughout and not a loose rubble-pile. The NIS measurements revealed a small variation in albedo across the surface of Eros after photometric corrections were made to correct for the differing viewing geometries (Bell et al. 2002; Clark et al. 2002). Brighter regions were thought to represent fresher and less altered material, whereas darker regions were thought to represent more space weathered material. The NLR range results were able to compare and contrast the roughness of smooth and rough regions visible in the images to the surfaces of the Moon and Mars. During the descent to the surface, NEAR Shoemaker obtained images with resolutions of up to 1 cm per pixel, revealing local differences in surface texture (Veverka et al. 2001). The landing area was generally blocky but with smooth regions in between, and, in some cases, the blocks were partially buried by finer regolith. There was a surprising lack of small craters visible in the descent images suggesting that they might have been filled in by regolith movement.

Unfortunately, there was no thermal-infrared instrument on NEAR Shoemaker but Eros was observed at thermal wavelengths in detail from Earth in 1974/1975 (Morrison 1976; Lebofsky & Rieke 1979), 1998 (Harris & Davies 1999; Müller 2007), and 2002 (Lim et al. 2005; Wolters et al. 2008). As the shape of Eros was not known at that time, Morrison (1976) corrected the thermal light-curve for shape effects using detailed optical light-curves to obtain an upper limit for the thermal inertia to be ~100 J m$^{-2}$ K$^{-1}$ s$^{-1/2}$, and a diameter of 22 ± 2 km at light-curve maximum. Lebofsky & Rieke (1979) fitted a crude shape model to the optical light-curves and used that in their thermophysical model of Eros. They explicitly included heat conduction, and thermal-infrared beaming was approximated in a simple way, to obtain a thermal inertia between 140 and 280

J m$^{-2}$ K$^{-1}$ s$^{-1/2}$, and a size of 39.3 x 16.1 x 16.1 km. Harris & Davies (1999), Lim et al. (2005), and Wolters et al. (2008) all used the NEATM thermal model of Harris (1998) to fit their data and obtained effective diameters that ranged from 20.4 to 23.6 km, and NEATM beaming parameters that ranged from 0.7 to 1.07. From their derived NEATM beaming parameter value, Harris & Davies (1999) indirectly concluded that the thermal inertia of Eros was ~170 J m$^{-2}$ K$^{-1}$ s$^{-1/2}$. These thermal inertia values are 2 to 5 times greater than that of the lunar surface, i.e. ~50 J m$^{-2}$ K$^{-1}$ s$^{-1/2}$ (Wesselink 1948), which is consistent with the regolith found and characterised by NEAR Shoemaker (Gundlach & Blum 2013).

The most recent study of the thermophysical properties of Eros has been performed by Müller (2007). In this case, the author combines the NEAR Shoemaker derived shape and spin state, i.e. that of Thomas et al. (2002), with a thermophysical model similar to that of Lagerros (1998) to analyse the data of Harris & Davies (1999). This dataset consists of 7 spectra taken at 7 different rotation phases of Eros (see Table 1 for observational geometry), and each contain simultaneous photometry at 25 wavelengths between 8 and 13 μm. Müller (2007) fits all 7 spectra simultaneously using the shape-based thermophysical model to derive a thermal inertia between 100 and 200 J m$^{-2}$ K$^{-1}$ s$^{-1/2}$, and a diameter of 17.8 ± 0.5 km. In this case, the lower thermal inertia bound comes from low surface roughness, and the upper bound from high surface roughness (e.g. see Figure 1). The derived diameter is ~5% larger than that found by NEAR Shoemaker, which is thought to be caused by the thermophysical model neglecting shadowing and self-heating effects within Eros's prominent concavities. Nevertheless, a thermal inertia of 150 ± 50 J m$^{-2}$ K$^{-1}$ s$^{-1/2}$ derived by Müller (2007) is the currently accepted value for Eros (e.g. Delbo et al. 2015).

The Lim et al. (2005) and Wolters et al. (2008) datasets also consist of spectra but were taken at a time when Eros was both illuminated and viewed almost pole-on (see Table 1 for observational geometry). In this geometry, thermal inertia has little influence on the observed thermal-infrared spectrum (e.g. see Figure 2), and so no thermophysical model fits were attempted in these previous works. However, surface roughness does have a large influence on the observed thermal-infrared spectrum (e.g. see Figure 3), but it was not considered at that time. In section 2 of this paper, the "almost pole-on" geometry of these previously unused datasets is utilised to constrain the surface roughness of Eros.

## 2. THERMOPHYSICAL MODELLING

### 2.1 The Advanced Thermophysical Model

The Advanced Thermophysical Model (or ATPM; Rozitis & Green 2011, 2012, 2013) is used to investigate the thermal-infrared beaming effect of Eros and to constrain its degree of surface roughness. The ATPM was developed to interpret thermal-infrared observations of airless planetary surfaces, and to simultaneously make asteroidal Yarkovsky and YORP effect predictions. It has been verified by accurately reproducing the lunar thermal-infrared beaming effect using properties measured in-situ by Apollo astronauts (Rozitis & Green 2011), and has been successfully applied to several near-Earth asteroids to determine their thermal and physical properties [e.g. (1620) Geographos in Rozitis & Green (2014), (1862) Apollo in Rozitis et al. (2013), (29075) 1950 DA in Rozitis et al. (2014), and (175706) 1996 FG3 in Wolters et al. (2011)].

To summarise how it works (Rozitis et al. 2013), the ATPM computes the surface temperature variation for each shape model facet during an asteroid rotation by solving one-dimensional heat conduction with a surface boundary condition that includes direct and multiply scattered sunlight, shadowing, and re-absorbed thermal radiation from interfacing facets (i.e. global self-heating effects). Rough-surface thermal-infrared beaming is explicitly included in the form of hemispherical craters, and the degree of roughness for each shape model facet is specified by the fraction of its area covered by these craters, $f_R$ (i.e. $f_R = 0$ for a fully smooth surface, and $f_R = 1$ for a fully rough surface). In terms of RMS slope, $\vartheta$, this can be calculated from the roughness fraction by using

$$\theta = 50\sqrt{f_R} \qquad (1)$$

for the hemispherical craters (Rozitis & Green 2011). The asteroid thermal emission as a function of wavelength, rotation phase, and various thermophysical properties is determined by applying the Planck function to the derived temperatures and summing across visible shape model facets.

### 2.2 Thermal-Infrared Flux Fitting

The Lim et al. (2005) dataset consists of 3 spectra taken at 3 different rotation phases (see Table 1), and the Wolters et al. (2008) dataset, which was obtained a week later, consists of 1 spectrum taken at a similar rotation phase to the 3rd spectrum of Lim et al. (2005). Both datasets are publicly available. The Lim et al. (2005) dataset contains 42 wavelengths between 8 and 13 μm per spectrum, and the Wolters et al. (2008) dataset contains 13 wavelengths between the same range. In this work, each spectrum is fitted independently by the ATPM to look for any spatial variations in surface roughness across Eros. In order to do this, the 7790 facet shape model of Eros is used (Thomas et al. 2002) along with its measured pole orientation (i.e. $\lambda = 17.2°$ and $\beta = 11.3°$ in ecliptic coordinates) and rotation period. The rotation period is known with sufficient accuracy to calculate the exact rotation phase Eros was observed at, and this information is used in the ATPM. Additionally, the measured disc-integrated Bond albedo of 0.12 (Domingue et al. 2002) is used as the model Bond albedo, and the mean diameter of Eros is kept fixed at the NEAR Shoemaker derived value of 16.92 km (Thomas et al. 2002). The thermal inertia is allowed to vary between 100 and 200 J m$^{-2}$ K$^{-1}$ s$^{-1/2}$ (Müller 2007) but as demonstrated in Figure 2 it has very little influence on the predicted flux and derived surface roughness. Therefore, thermal inertia can be considered to be a fixed parameter in the ATPM modelling and fitting.

Since thermal-infrared spectra have potentially large uncertainties in their absolute flux calibration, which would usually map to additional diameter and albedo uncertainty in the model fitting, an additional instrument scale factor, *ISF*, parameter is included. This parameter simply scales the model flux up or down to take into account any systematic offsets in the spectra caused by their absolute flux calibration. This means that the surface roughness is derived by fitting to the colour temperatures exhibited by the thermal spectra rather than fitting to their absolute flux values. Lim et al. (2005) estimate the uncertainty of their absolute flux calibration to be 10% whilst Wolters et al. (2008) estimate it to be 7% for their calibration. Therefore, the instrument scale factor is allowed to vary between 0.5 and 1.5 to take into account these uncertainties.

The free parameters to be constrained by fitting the ATPM to each spectrum are therefore the roughness fraction, $f_R$, and the instrument scale factor, *ISF*. This is performed by minimising the $\chi^2$ fit between the model thermal flux predictions, $F_{MOD}(\lambda_n, f_R)$, the observations, $F_{OBS}(\lambda_n)$, and the observational errors, $\sigma_{OBS}(\lambda_n)$, by using

$$\chi^2 = \sum_{n=1}^{N} \left[ \frac{ISF \cdot F_{MOD}(\lambda_n, f_R) - F_{OBS}(\lambda_n)}{\sigma_{OBS}(\lambda_n)} \right]^2 \qquad (2)$$

for a set of $n = 1$ to $N$ observations with wavelength $\lambda_n$. The roughness fraction and instrument scale factor are varied through their plausible ranges in steps of 0.01 to form a 2-dimensional grid of model test parameters in order to find the minimum $\chi^2$. A region bounded by a constant $\Delta\chi^2$ at the 1-$\sigma$ confidence level around the minimum $\chi^2$ then gives the uncertainty on the derived parameters (i.e. $\Delta\chi^2 = 2.3$ for 2 free parameters at 1-$\sigma$ confidence).

## 2.3 Results

The ATPM fits to the 3 spectra of Lim et al. (2005) give surface roughness RMS slopes of 40.0 $^{+8.0}/_{-8.8}$, 37.7 $^{+6.4}/_{-10.4}$, and 35.0 $^{+9.7}/_{-5.4}$ degrees, and instrument scale factors of 0.99 $^{+0.06}/_{-0.07}$, 1.04 $^{+0.06}/_{-0.09}$, and 1.00 $^{+0.06}/_{-0.05}$ respectively. The fit to the Wolters et al. (2008) spectrum gives an RMS slope of 19.4 $^{+1.3}/_{-5.2}$ ° and an instrument scale factor of 1.30 $^{+0.02}/_{-0.03}$. Figure 3 demonstrates the fits to these datasets, and the results are summarised in Table 1.

The ATPM fits to the Lim et al. (2005) spectra are consistent with one another, and the instrument scale factors range from 0.92 to 1.1, which are also consistent with the 10% uncertainty on the absolute flux calibration. However, the fit to the Wolters et al. (2008) dataset produces a much smaller degree of surface roughness, which is apparent by the cooler colour temperature of the dataset's thermal spectrum. This is surprising considering that this thermal spectrum was taken at very similar geometry and rotation phase to the 3rd spectrum of Lim et al. (2005). The derived instrument scale factor of 1.3 is also very high, and cannot be explained completely by the 7% uncertainty on the absolute flux calibration. In fact, the difference in instrument scale factors between the two datasets explain the difference in diameters obtained by the previously published NEATM fitting, i.e. 20.4 km for Lim et al. (2005) and 23.3 km for Wolters et al. (2008). This suggests that the Wolters et al. (2008) observations could have been affected by external environmental factors and could be anomalous. Therefore, the fit results of the Lim et al. (2005) dataset are considered to be the most reliable surface roughness measurements of Eros, and will be used for comparison purposes in section 3. Nevertheless, the fit to the Wolters et al. (2008) dataset still demonstrates how surface roughness can be extracted using this "almost pole-on" geometry.

## 3. DISCUSSION

### 3.1 Geological Interpretation of Surface Roughness Derived by Thermal-Infrared Beaming

The mean RMS slope of 38 ± 8° derived from the Lim et al. (2005) datasets is similar to the ~32° derived for the lunar surface by Rozitis & Green (2011). The spatial scales at which this surface roughness statistic is relevant start at the thermal skin depth and range up to the facet size of the shape model used (i.e. ~540 m). The thermal skin depth, $l$, is given by

$$l = \frac{\Gamma}{\rho C} \sqrt{\frac{P}{2\pi}} \qquad (3)$$

where $\Gamma$ is the thermal inertia, $\rho$ is the density, and $C$ is the specific heat capacity of the surface, and $P$ is the period of interest (i.e. rotation period for the diurnal skin depth, and orbital period for the seasonal skin depth). By assuming a density of 3150 kg m$^{-3}$ and a specific heat capacity of 550 J kg$^{-1}$ K$^{-1}$, which are typical values for ordinary chondrites (Opeil et al. 2010), equation (3) gives a diurnal thermal skin depth of ~0.5 cm when using Eros's thermal inertia and rotation period of 150 J m$^{-2}$ K$^{-1}$ s$^{-1/2}$ and 5.27 hr, respectively. The corresponding seasonal thermal skin depth is ~0.3 m for an orbital period of 1.76 yr. Given that the inferred degree of surface roughness is rather large then it most likely corresponds to the diurnal skin depth scale. Temperature gradients at this scale are still possible since the sub-solar latitude was -68°, which would still cause small diurnal temperature variations on the sunlit side, especially near the terminator. In fact, most of the flux enhancement caused by thermal-infrared beaming come from surfaces on the limb anyway, as demonstrated in Figure 4.

It is important for validation purposes that this surface roughness is compared with that measured by other means. The highest resolution shape model of Eros has a spatial scale of ~27 m with ~3 million facets (Gaskell 2008), which is too large for direct comparison purposes despite the huge number of resolved facets. This shape model gives an RMS slope of 10.5 ± 4.2 degrees at ~27 m scales when its facet orientations are compared against those of the 7790 facet shape model used in the ATPM.

Fortunately, and as mentioned previously, the laser range finder (NLR) flown on the NEAR Shoemaker mission obtained more than 16 million range returns from Eros (Cheng *et al.* 2002). The NLR determined the range to within a resolution of 0.312 m but due to uncertainty of the spacecraft ephemerides the effective resolution was <6 m. However, over time spans <1 hour in duration the ephemeris errors could be ignored, as they varied slowly, to study small-scale topography with a precision of ~1 m. After subtracting the local radius from the NLR profiles the surface roughness statistics could be studied. These statistics included the standard deviation in height differences, $v(\Delta x)$, for a given spatial scale $\Delta x$. In a self-affine structure, the standard deviation in height differences obey

$$v(\Delta x) = v_0 (\Delta x)^H \qquad (4)$$

where $H$ is the Hurst exponent and $v_0$ is a normalising constant. These quantities allow the RMS slope for a given spatial scale, $\vartheta(\Delta x)$, to be calculated from

$$\theta(\Delta x) = \tan^{-1}\left(\frac{\nu(\Delta x)}{\Delta x}\right). \qquad (5)$$

For Eros, these roughness statistics were determined for 6 specific tracks on the asteroid for spatial scales between 1 and 1000 m (Cheng et al. 2002), and these results are shown and summarised in Figure 5 and Table 2 respectively. These results demonstrated that Eros has a fractal-like structure over spatial scales ranging from a few metres to hundreds of metres. The fractal statistics indicated that the surface of Eros is very rough, especially at large scales. They are also similar for diverse regions of Eros, which indicated that similar surface shaping processes, driven by collisions and gravity, occur at the scales observed in these regions.

Although not directly measured, the surface roughness at 0.5-cm scales can be estimated by extrapolating the NLR results when assuming that the Hurst exponent remains constant from metre to millimetre spatial scales. Extrapolating the results of the 6 tracks on Eros to a 0.5-cm spatial scale gives RMS slopes that range from 20.0 to 29.4 degrees (see Figure 5 and Table 2), and these are slightly less than the RMS slope of 38 ± 8° measured by the thermal-infrared beaming effect. This either implies that additional surface shaping processes could be making the surface of Eros rougher at spatial scales smaller than one metre, or that it is a result of assuming a fractional coverage of hemispherical craters in the thermal-infrared beaming model. In particular on Eros, there is a marked deficiency of craters at spatial scales under 10 m such that the surface roughness is actually dominated by boulders (Chapman 2002). However, detailed thermophysical modelling of more realistic surface roughness topography has demonstrated that the hemispherical crater produces comparable degrees of beaming for the same degree of surface roughness present in the topography models investigated (Rozitis & Green 2011; Davidsson et al. 2015). Therefore, the slight elevation of surface roughness at small spatial scales implied here on Eros could be real.

Elevated surface roughness at small spatial scales has previously been implied for other near-Earth asteroids whose surfaces have been probed by radar. In particular, the radar circular polarisation ratio is sensitive to surface roughness at spatial scales that range upwards from the radar wavelength (Ostro 1993). For near-Earth asteroids (4179) Toutatis and (25143) Itokawa (Nolan et al. 2013), their 3.5-cm circular polarisation ratios (i.e. 0.29 ± 0.01 and 0.47 ± 0.04 respectively) are much higher than their 12.5-cm circular polarisation ratios (i.e. 0.23 ± 0.03 and 0.26 ± 0.04 respectively), which implies that their surfaces are rougher at the 3.5-cm scale. Unfortunately for Eros, the uncertainties of its 3.5-cm and 12.5-cm circular polarisation ratios, i.e. 0.33 ± 0.07 and 0.28 ± 0.06, are too large to confirm or deny whether its surface is rougher at the 3.5-cm scale. Nevertheless, the additional physical processes that could make the surface of Eros rougher at spatial scales smaller than one metre include thermal fracturing (Delbo et al. 2014), dust levitation (Hughes et al. 2008), and possibly many others [e.g. see review by Murdoch et al. (2015)].

### 3.2 Application to Other Asteroids

The "pole-on" illumination geometry could be exploited to investigate the surface roughness of other high obliquity asteroids. To determine which illumination sub-solar latitudes and observation phase angles would be useful, a small parameter study is performed. In this study, a 1-km spherical asteroid with an S-type albedo (i.e. a geometric albedo of 0.25), a typical 6 hour rotation period, and a typical near-Earth asteroid thermal inertia of 200 J m$^{-2}$ K$^{-1}$ s$^{-1/2}$ (Delbo et al. 2007) is placed 1.3 AU from the Sun. It is then observed from a distance of 0.3 AU with sub-solar latitude varying from 0 to 90 degrees, and with phase angle varying from 0 to 180 degrees.

Figure 6 shows the disc-integrated flux enhancement caused by rough surface thermal-infrared beaming as a function of phase angle for different wavelengths at a fixed sub-solar latitude of 90° (i.e. exactly "pole-on" illumination). As demonstrated, the flux enhancement is greatest for phase angles <40°, and the flux is actually reduced for phase angles >70°. Therefore, these phase angle ranges would be most suitable for characterising the surface roughness during "pole-on" illumination geometries. Applying the NEATM thermal model (Harris 1998) to the flux enhancements shown in Figure 6 gives the variation of the NEATM beaming parameter with phase angle for this "pole-on" asteroid, which is shown in Figure 7. For comparison purposes, the empirical linear relationship of the NEATM beaming parameter, $\eta$, with phase angle, $\alpha$, determined by Wolters et al. (2008) for a range of near-Earth asteroids is also plotted:

$$\eta = \left(0.91 \pm 0.17\right) + \left(0.013 \pm 0.004\right)\alpha .$$ (6)

As indicated, the NEATM beaming parameter for the "pole-on" asteroid is lower than the typical value expected for a near-Earth asteroid at all phase angles up to 100°. To confirm this for Eros, the NEATM beaming parameter value of 1.07 ± 0.21 obtained during "equatorial" illumination (Harris & Davies 1999) falls within the typical near-Earth asteroid range, and the value of 0.71 ± 0.08 obtained during "almost pole-on" illumination (Lim et al. 2005) falls on the "pole-on" asteroid trend.

To assess which sub-solar latitudes surface roughness effects would become dominant, the relative influences of thermal inertia and surface roughness on the predicted flux emitted by an asteroid as a function of sub-solar latitude are compared. For thermal inertia effects, the difference in predicted emitted flux between thermal inertia values of 130 and 270 J m$^{-2}$ K$^{-1}$ s$^{-1/2}$ are compared to a nominal emitted flux prediction for a thermal inertia of 200 J m$^{-2}$ K$^{-1}$ s$^{-1/2}$ such that the relative influence of thermal inertia, $\Gamma_{INF}$, is given by

$$\Gamma_{\mathrm{INF}} = \frac{\left| F_{\mathrm{MOD}}\left(\Gamma = 270, f_{\mathrm{R}} = 0.5\right) - F_{\mathrm{MOD}}\left(\Gamma = 130, f_{\mathrm{R}} = 0.5\right)\right|}{F_{\mathrm{MOD}}\left(\Gamma = 200, f_{\mathrm{R}} = 0.5\right)} .$$ (7)

This thermal inertia range was chosen because it is equivalent to the ~33% uncertainty on Eros's measured thermal inertia value. A moderate level of surface roughness is assumed in this thermal inertia test. Similarly, surface roughness effects are assessed by the difference in predicted emitted flux between extremely rough and smooth surfaces compared to that of a moderately rough surface. The relative influence of surface roughness, $\vartheta_{INF}$, is then given by

$$\theta_{\mathrm{INF}} = \frac{\left| F_{\mathrm{MOD}}\left(\Gamma = 200, f_{\mathrm{R}} = 1.0\right) - F_{\mathrm{MOD}}\left(\Gamma = 200, f_{\mathrm{R}} = 0.0\right)\right|}{F_{\mathrm{MOD}}\left(\Gamma = 200, f_{\mathrm{R}} = 0.5\right)} ,$$ (8)

where a nominal thermal inertia value is assumed in this case. The relative influences of thermal inertia and surface roughness are then assessed as a function of sub-solar latitude using equations (7) and (8) at a chosen wavelength of 8 μm and averaged over phase angles between 0 and 40°. A wavelength of 8 μm was chosen in these calculations because usually it is the shortest wavelength observed in the mid-infrared spectral region, and is therefore the most sensitive wavelength to temperature changes caused by thermal inertia and surface roughness. Longer wavelengths are less sensitive to these temperature changes, and therefore the relative influences of thermal inertia and surface roughness both decrease slightly with increasing wavelength. The results for a wavelength of 8 µm are shown in Figure 8 for the example 1-km spherical asteroid considered here, and for Eros during "almost pole-on" illumination conditions.

For sub-solar latitudes of less than ~40°, thermal inertia and surface roughness have relatively equal influences on the predicted flux emitted by an asteroid, which leads to degeneracy between the two parameters. For sub-solar latitudes greater than ~60°, the influence by thermal inertia has dropped to very little whilst the influence by surface roughness has been enhanced. This allows surface roughness to be constrained during high sub-solar latitude illumination conditions, and explains why the surface roughness of Eros could be constrained using the Lim et al. (2005) observations. Therefore, a dedicated observational campaign with a thermal-infrared instrument could reveal new insights into the rough surface beaming effect for high obliquity asteroids when they are illuminated with sub-solar latitudes greater than 60°, and are observed at phase angles of less than 40°.

## 4. SUMMARY AND CONCLUSIONS

In summary, the "almost pole-on" illumination and viewing geometry of Eros during the September 2002 observations (Lim et al. 2005; Wolters et al. 2008) produced thermal flux emissions that depended primarily on surface roughness. The thermal-infrared beaming effect of Eros was accurately modelled with the Advanced Thermophysical Model (Rozitis & Green 2011, 2012, 2013) by using the exact shape model and physical properties measured in-situ by the NEAR Shoemaker mission (Thomas et al. 2002). This allowed a surface roughness RMS slope of 38 ± 8° to be measured for spatial scales ranging upwards from the diurnal thermal skin depth, i.e. ~0.5 cm for Eros. This value is slightly greater than the RMS slope of 25 ± 5° estimated from the NEAR Shoemaker laser ranging results (Cheng et al. 2002) when extrapolated to this spatial scale. This indicates that other surface shaping processes [e.g. see review by Murdoch et al. (2015)] might operate, in addition to collisions and gravity, at spatial scales smaller than one metre in order to make asteroid surfaces rougher. Surface roughness on other high obliquity asteroids can be constrained during "pole-on" illumination conditions when the sub-solar latitude is greater than 60°, and if they are observed at phase angles of less than 40°. Furthermore, it is predicted that they would exhibit NEATM beaming parameters that are lower than expected for a typical asteroid at all phase angles up to 100°.

In conclusion, these results have important implications for breaking the degeneracy between thermal inertia and surface roughness seen in disc-integrated thermal observations of asteroids, and for the geological interpretation of surface roughness measured by beaming. In particular, the upcoming asteroid sample return missions OSIRIS-REx (Lauretta et al. 2015) and Hayabusa 2 (Okada et al. 2012) both have thermal-infrared instruments onboard to aid sample site selection, and therefore will be capable of investigating rough surface thermal-infrared beaming in-situ at their target asteroids.

### Acknowledgements

BR acknowledges support from the Royal Astronomical Society (RAS) in the form of a research fellowship. BR is also grateful of the anonymous referee for constructive comments that helped to improve the manuscript.

**Figures**

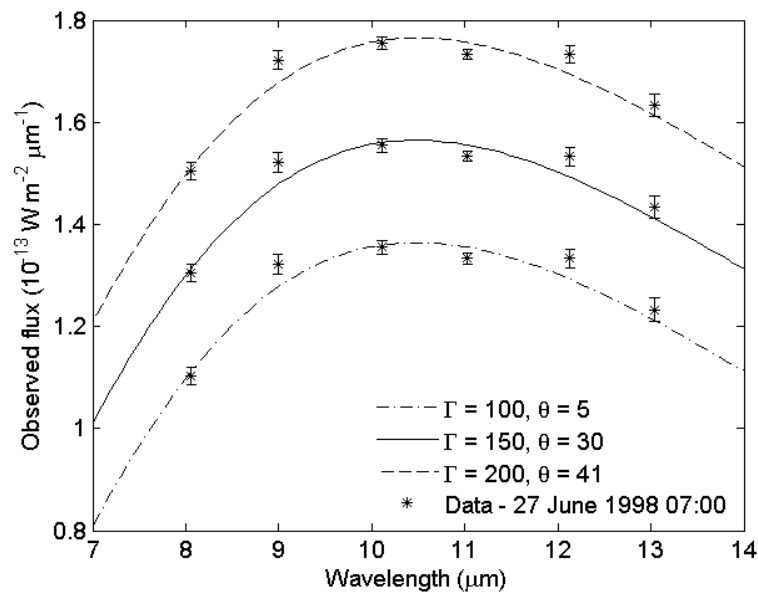

**Figure 1:** Degeneracy between thermal inertia and surface roughness seen in an example thermal-infrared spectrum of (433) Eros taken during "equatorial" solar illumination. The spectrum shown here (data points) is a subset of the data collected by Harris & Davies (1999), and is representative of Eros at light-curve maximum on 27 June 1998 07:00 (see Table 1 for observational geometry). The error bars correspond to the 1-$\sigma$ uncertainties on the measured data points. Since the ATPM produces almost identical thermal spectra for the three different combinations of thermal inertia and surface roughness (lines), the model spectra are offset for clarity (i.e. by -0.2 x $10^{-13}$ W m$^{-2}$ μm$^{-1}$ for a thermal inertia of 100 J m$^{-2}$ K$^{-1}$ s$^{-1/2}$, and by +0.2 x $10^{-13}$ W m$^{-2}$ μm$^{-1}$ for a thermal inertia of 200 J m$^{-2}$ K$^{-1}$ s$^{-1/2}$). As demonstrated, the lower thermal inertia value fits well with low surface roughness, and the higher thermal inertia value fits equally well but with high surface roughness.

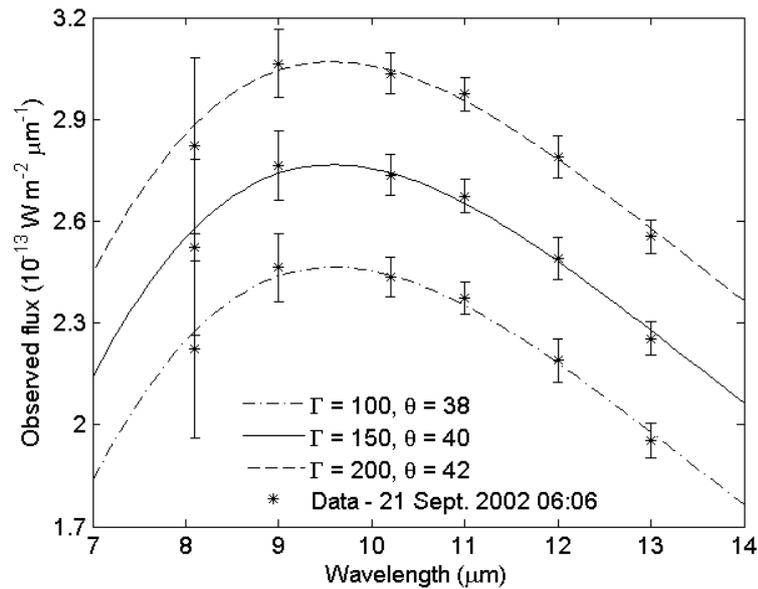

**Figure 2:** Absent degeneracy between thermal inertia and surface roughness seen in an example thermal-infrared spectrum of (433) Eros when illuminated and viewed "almost pole-on". The spectrum shown here (data points) is a subset of the data collected by Lim et al. (2005) on 21 September 2002 06:06 (see Table 1 for observational geometry). The error bars correspond to the 1-$\sigma$ uncertainties on the measured data points. Since the ATPM produces almost identical thermal spectra for the three different combinations of thermal inertia and surface roughness (lines), the model spectra are offset for clarity (i.e. by -0.3 x $10^{-13}$ W m$^{-2}$ μm$^{-1}$ for a thermal inertia of 100 J m$^{-2}$ K$^{-1}$ s$^{-1/2}$, and by +0.3 x $10^{-13}$ W m$^{-2}$ μm$^{-1}$ for a thermal inertia of 200 J m$^{-2}$ K$^{-1}$ s$^{-1/2}$). As demonstrated, the fitted degree of surface roughness is almost independent of the thermal inertia value.

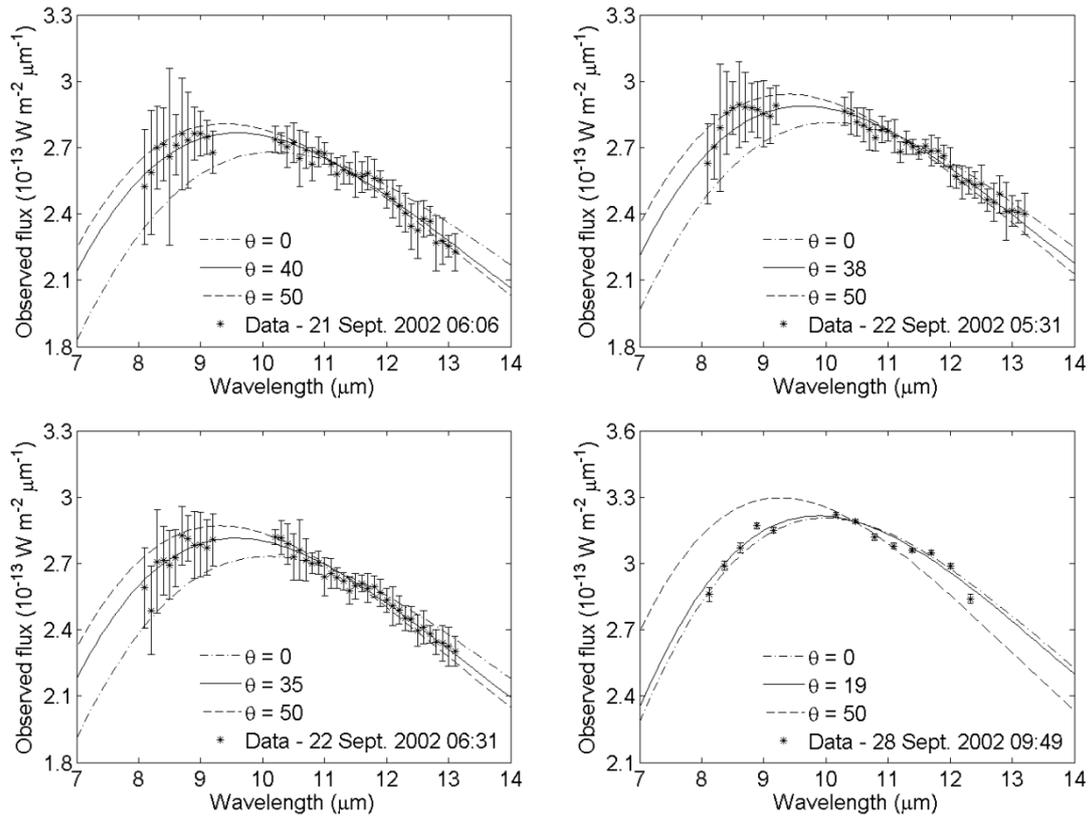

**Figure 3:** Surface roughness derivation of (433) Eros by fitting the ATPM (lines) to "almost pole-on" thermal-infrared spectra (data points) taken in September 2002. The error bars correspond to the 1-$\sigma$ uncertainties on the measured data points. The Lim et al. (2005) dataset (first three panels) consists of 42 wavelengths (or data points) per spectrum and had best $\chi^2$ fits of 6.4, 9.9, and 6.5, as obtained by the thermal-infrared flux fitting described in section 2.2 using equation (2). Likewise, the Wolters et al. (2008) dataset (bottom right panel) consisted of 13 wavelengths for its spectrum and had a best $\chi^2$ fit of 67.2.

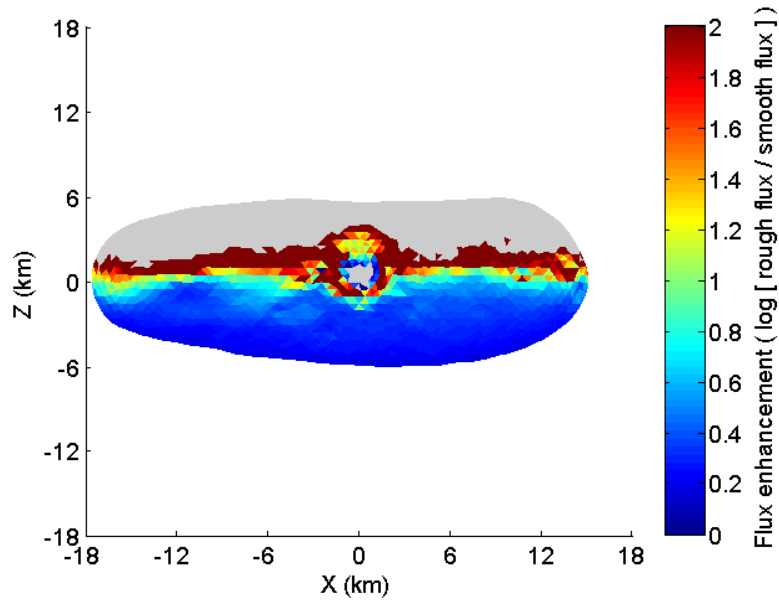

**Figure 4:** 8 μm thermal-infrared beaming map of (433) Eros for the Lim et al. (2005) geometry (see Table 1). The 8 μm flux enhancement shown here is calculated by rotationally averaging the rough and smooth flux emissions observed for each shape model facet and then taking their ratio. The logarithm of this ratio is plotted here to better highlight the variation in enhancement caused by beaming across the surface of Eros. In this geometry, Eros is illuminated and viewed approximately along the z-axis from below, and hence explains why the northern hemisphere of Eros makes no contribution to the observed flux (greyed out region). As demonstrated, the flux enhancement is greatest near the equatorial region where the terminator lies in this "almost pole-on" illumination geometry, and is caused by the limb brightening effect described in Rozitis & Green (2011).

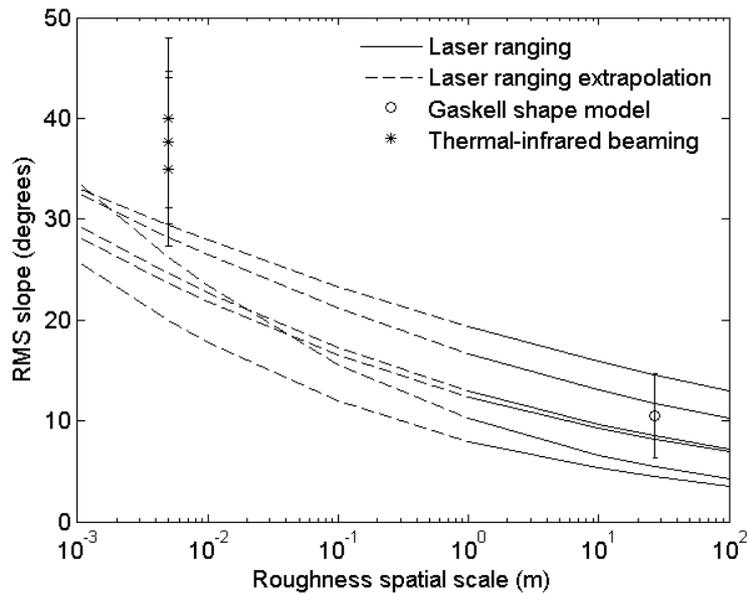

**Figure 5:** Surface roughness RMS slope of (433) Eros as a function of spatial scale. The solid lines represent the surface roughness measured by NEAR Shoemaker using laser ranging for the 6 different tracks listed in Table 2 (Cheng et al. 2002). The dashed lines represent the extrapolation of the laser ranging results to millimetre spatial scales assuming that the measured Hurst exponents remain constant. The open circle and error bar represent the mean and standard deviation of the RMS slopes measured when the ~3 million facet shape model of Gaskell (2008) is compared against the 7790 facet shape model of Thomas et al. (2002). The asterisks represent the surface roughness derived by fitting the ATPM to the Lim et al. (2005) spectra via the thermal-infrared beaming effect (see Table 1), and their error bars correspond to the 1-$\sigma$ uncertainties on the derived data points.

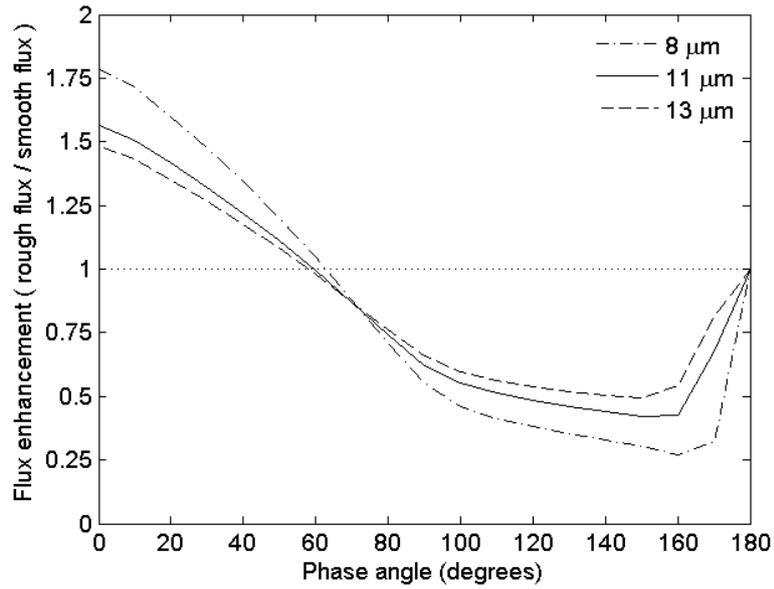

**Figure 6:** Disc-integrated flux enhancement by rough surface thermal-infrared beaming as a function of phase angle and wavelength for a "pole-on" asteroid. This example is for a 1-km spherical asteroid placed at a heliocentric distance of 1.3 AU, and has a rotation pole pointed directly at the Sun.

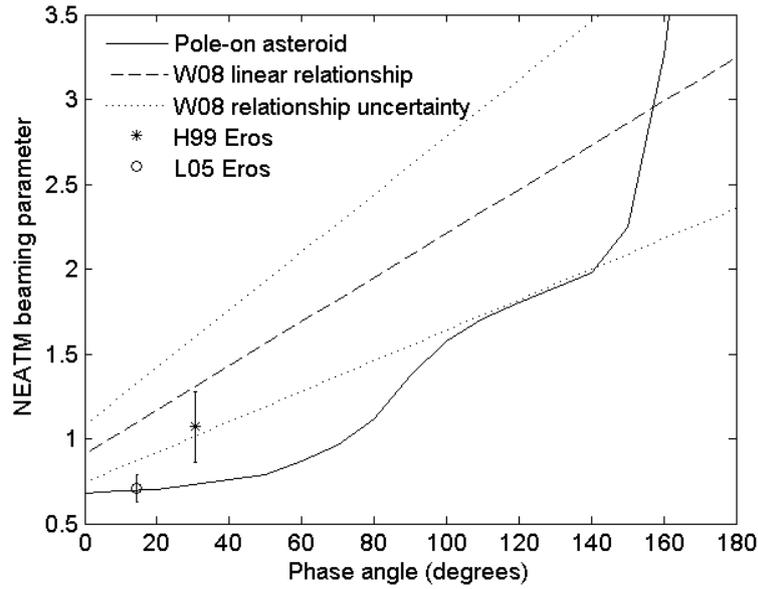

**Figure 7:** Variation of the NEATM beaming parameter as a function of phase angle for a "pole-on" asteroid. The NEATM beaming parameter values have been calculated from the disc-integrated flux enhancements shown in Figure 6 for the same example asteroid. For comparison purposes, the linear relationship of the NEATM beaming parameter with phase angle determined by Wolters et al. (2008) for a range of near-Earth asteroids is also plotted (W08). Additionally, the NEATM beaming parameters determined for (433) Eros during "equatorial" (H99 - Harris & Davies 1999) and "almost pole-on" (L05 - Lim et al. 2005) illumination conditions are also identified.

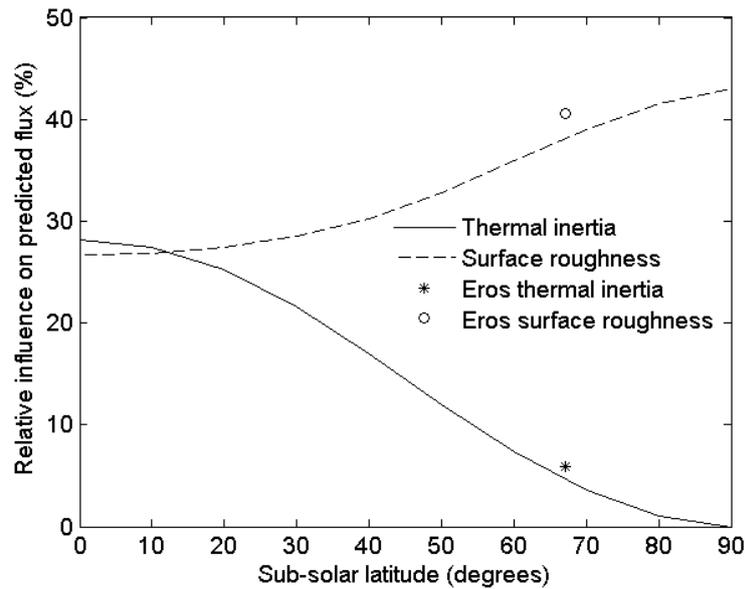

**Figure 8:** Relative influence of thermal inertia and surface roughness on the predicted flux emitted by an asteroid as a function of sub-solar latitude. The relative influences of thermal inertia and surface roughness are calculated using equations (7) and (8) for the same example asteroid used in Figures 6 and 7 (lines). For comparison purposes, the equivalent values calculated for (433) Eros during "almost pole-on" illumination conditions are also plotted (data points).

**Tables**

| Dataset | UT Date | Telescope | r (AU) | Δ (AU) | α (deg) | Sub-solar latitude (deg) | Sub-observer latitude (deg) | Rotation phase (deg) | Surface roughness RMS slope (deg) |
|---------|---------|-----------|--------|--------|---------|--------------------------|-----------------------------|----------------------|-----------------------------------|
| Harris & Davies (1999) | 1998 Jun. 27 07:00 | UKIRT | 1.617 | 0.796 | 30.6 | 34.6 | 64.0 | 235 | No constraint |
| Lim et al. (2005) | 2002 Sep. 21 05:48 to 06:23 | Palomar | 1.608 | 0.637 | 14.5 | -67.1 | -60.1 | 89 to 129 | $40.0\ ^{+8.0}/_{-8.8}$ |
| | 2002 Sep. 22 05:16 to 05:46 | Palomar | 1.606 | 0.637 | 15.0 | -67.6 | -59.6 | 252 to 286 | $37.7\ ^{+6.4}/_{-10.4}$ |
| | 2002 Sep. 22 06:16 to 06:46 | Palomar | 1.606 | 0.637 | 15.0 | -67.6 | -59.7 | 321 to 355 | $35.0\ ^{+9.7}/_{-5.4}$ |
| Wolters et al. (2008) | 2002 Sep. 28 09:39 to 09:58 | UKIRT | 1.589 | 0.640 | 18.2 | -70.4 | -57.1 | 307 to 329 | $19.4\ ^{+1.3}/_{-5.2}$ |

**Table 1:** Observation geometry and surface roughness fitting results for the 3 different datasets of (433) Eros.

| Track on (433) Eros | Hurst Exponent, $H$ | RMS of Height Differences, $v_0$ (m) | Extrapolated RMS Slope, $\vartheta$, for 0.5 cm (deg) |
|---------------------|---------------------|--------------------------------------|------------------------------------------------------|
| Himeros | 0.81 | 0.18 | 26.2 |
| Twist | 0.89 | 0.30 | 28.3 |
| Groove | 0.82 | 0.14 | 20.0 |
| Psyche Rim | 0.91 | 0.35 | 29.4 |
| Psyche Debris | 0.87 | 0.22 | 23.7 |
| Psyche Wall | 0.87 | 0.23 | 24.6 |

**Table 2:** Summary of surface roughness statistics determined for 6 laser ranging tracks on (433) Eros [modified from Cheng et al. (2002)].